\documentclass{llncs}
\usepackage[dvips]{graphicx}
\usepackage[latin1]{inputenc}
\usepackage{amssymb,amsmath,array}
\usepackage[USenglish]{babel}

\usepackage{url}
\usepackage{mathtools}
\DeclarePairedDelimiter{\ceil}{\lceil}{\rceil}
\usepackage{xcolor}
\usepackage{amsmath}
\usepackage{subfigure}
\usepackage{algorithm}
\usepackage{algorithmic}
\usepackage{wrapfig}
\usepackage[font=small]{caption}

\usepackage{amssymb}
\usepackage{xspace}

\newcommand{\etal}{\emph{et al.}\xspace}
\usepackage{tikz}
\usetikzlibrary{shapes,arrows,shadows}
\usetikzlibrary{mindmap,trees}
\definecolor{mygreen}{rgb}{0.553,0.682,0.063}
\definecolor{myblue}{rgb}{0.0,0.208,0.376}
\definecolor{mygray}{rgb}{0.906,0.906,0.906}

\newcommand{\tdim}{d}

\newcommand{\pointset}{P}
\newcommand{\queryset}{Q}
\newcommand{\tsize}{n}
\newcommand{\testsize}{m}
\newcommand{\numleafchunks}{N}
\renewcommand{\vec}[1]{\mathbf{#1}}
\newcommand{\Reals}{{\mathbb R}}

\newcommand{\kdtree}{\mbox{\ensuremath{k}-d}~tree\xspace}

\newcommand{\kdtrees}{\mbox{\ensuremath{k}-d}~trees\xspace}

\newcommand{\KdTrees}{\mbox{\ensuremath{K}-d}~Trees\xspace}
\newcommand{\Kdtrees}{\mbox{\ensuremath{K}-d}~trees\xspace}
\def\O{{\cal O}}
\def\O{{\cal O}}
\newcommand{\ie}{i.\nolinebreak[4]\hspace{0.01em}\nolinebreak[4]e.\@\xspace}
\newcommand{\eg}{e.\nolinebreak[4]\hspace{0.01em}\nolinebreak[4]g.\@\xspace}

\newcommand{\GPU}{{\sc GPU}}
\newcommand{\CPU}{{\sc CPU}}

\newcommand{\GPUs}{{\sc GPUs}}

\newcommand{\BufferkdTrees}{\mbox{Buffer~\ensuremath{k}-d}~Trees\xspace}
\newcommand{\bufferkdtree}{\mbox{buffer~\ensuremath{k}-d}~tree\xspace}
\newcommand{\bufferkdtrees}{\mbox{buffer~\ensuremath{k}-d}~trees\xspace}

\begin{document}
%style file for ESANN manuscripts
\title{Bigger Buffer $k$-d Trees\\on Multi-Many-Core Systems}

%***********************************************************************
% AUTHORS INFORMATION AREA
%***********************************************************************
\author{Fabian Gieseke$^1$, Cosmin Eugen Oancea$^2$, Ashish Mahabal$^3$,\\Christian Igel$^2$, and Tom Heskes$^1$
%
% Optional short acknowledgment: remove next line if non-needed
% \thanks{This is an optional funding source acknowledgement.}
%
% DO NOT MODIFY THE FOLLOWING '\vspace' ARGUMENT
% \vspace{.3cm}\\
%
%
% Remove the next three lines in case of a single institution
% \vspace{.1cm}\\
% 2- Heidelberg Institute for Theoretical Studies gGmbH - Astroinformatics\\
% Schlo{\ss}-Wolfsbrunnenweg 35, 69118 Heidelberg - Germany
}

\institute{
% Addresses and institutions (remove "1- " in case of a single institution)
1 - Institute for Computing and Information Sciences - Radboud University Nijmegen\\
Toernooiveld 212, 6525 EC Nijmegen - The Netherlands\\
\{\url{f.gieseke,t.heskes}\}\url{@cs.ru.nl}\\
2 - Department of Computer Science - University of Copenhagen\\
Universitetsparken 5, 2100 Copenhagen - Denmark\\
\{\url{cosmin.oancea,igel}\}\url{@di.ku.dk}\\
3 - Caltech Astronomy - Caltech\\
1200 East California Blvd, Pasadena CA 91125 - USA\\
\url{aam@astro.caltech.edu}
}

%***********************************************************************
% END OF AUTHORS INFORMATION AREA
%***********************************************************************

\maketitle
\thispagestyle{plain}

\begin{abstract}
A \bufferkdtree~is a \kdtree~variant for massively-parallel nearest neighbor search. While providing valuable speed-ups on modern many-core devices in case both a large number of reference and query points are given, \bufferkdtrees are limited by the amount of points that can fit on a single device. In this work, we show how to modify the original data structure and the associated workflow to make the overall approach capable of dealing with massive data sets. We further provide a simple yet efficient way of using multiple devices given in a single workstation. The applicability of the modified framework is demonstrated in the context of astronomy, a field that is faced with huge amounts of data.
\end{abstract}

\section{Motivation}
\label{section:motivation}
Nearest neighbor search is a fundamental problem and an ingredient of
many state-of-the-art data analysis techniques. While being a
conceptually very simple task, the induced computations can quickly
become a major bottleneck in the overall workflow when both a large reference and a large query set are given. In the literature, many techniques have been proposed that aim at accelerating the search. Typical are include the use of spatial search structures, approximation schemes, and parallel implementations~\cite{Bentley1975,BeygelzimerKL2006,Cayton2012,GarciaDNB2010,IndykM1998,PanM2011}. A recent trend in the field of big data analytics is the application of massively-parallel devices such as \emph{graphics processing units} (\GPUs) to speed up the involved computations. While such modern many-core devices can significantly reduce the practical runtime, obtaining speed-ups over standard \CPU-based execution is often not straightforward and usually requires a careful adaptation of the sequential implementations.

Spatial search structures such as \kdtrees are an established way to reduce the computational requirements induced by nearest neighbor search for spaces of moderate dimensionality (\eg, up to $\tdim=30$). A typical parallel \kdtree based search assigns one thread to each query and all threads process the same tree \emph{simultaneously}. Such an approach, however, is not suited for \GPUs~since each thread might induce a completely different tree traversal, which results in massive branch divergence and irregular accesses to the device's memory. 

Recently, we have proposed a modification of the classical \kdtree data structure, called \emph{\bufferkdtree}, which aims at combining the benefits of both spatial search structures and massively-parallel devices~\cite{GiesekeHOI2014}. The key idea is to assign an additional buffer to each leaf of the tree and to \emph{delay} the processing of the queries reaching a leaf until enough work has been gathered. In that case, all queries stored in all buffers are processed together in a brute-force manner via the many-core device.
% using the \GPU, where all queries that are assigned to a particular buffer are compared with all patterns stored in the associated leaf (see below).
While the framework achieves significant speed-ups on modern many-core devices over both a massively-parallel brute-force execution on \GPUs~as well as over a multi-threaded \kdtree based search running on multi-core systems, it is limited by the amount of reference and query points that fit on a \GPU. In this work, we show how to remove this limitation by modifying the induced workflow to efficiently support huge reference and query point sets that are too large to be completely stored on the devices. This crucial modification renders \bufferkdtrees capable of dealing with huge data sets.

% The work is structured as follows: In Section~\ref{sec:background}, the background related to massively-parallel computing, classical \kdtree-based nearest neighbor search, and the recent \bufferkdtree~extension will be provided. Section~\ref{sec:processing_bigger_trees} describes the modifications made to render the framework scalable to huge data sets. The outcome of the experimental evaluation is given in Section~\ref{sec:experiments}, followed by conclusions drawn in Section~\ref{sec:conclusions}.

\section{Background}
\label{sec:background}
For the sake of completeness, we provide the background related to massively-parallel programming on \GPUs~as well as to classical \kdtree-based nearest neighbor search. We further sketch the key ideas of the \bufferkdtree extension.
% ; for details, we refer to our previous work~\cite{GiesekeHOI2014}.

\subsection{Architecture and Programming Model}
\label{subsec:GPUarch}

Modern many-core devices such as \GPUs~offer massive parallelism and can nowadays also be used for so-called \emph{general-purpose computations} such as matrix-matrix multiplication. In contrast to standard \CPU-based systems, \GPUs~rely on simplified control units and on a memory subsystem that does not attempt to provide the illusion of a uniform access cost to memory.  
%\cite{ChengGM2014}. 

\GPU~architectures are typically formed from a number of vector processors, several special function units and an amount of fast memory that is split between registers, L1 and L2 data caches, scratchpad and read-only memory.  Each vector processor consists of multiple execution units that execute in lock step, \ie, in a single-instruction multiple data ({\sc simd}) fashion, and each execution unit is further multi-threaded in order to hide memory latency.

The GPU programming model typically reflects this hardware organization: For example, expressing a parallel computation in \texttt{OpenCL}~\cite{Scarpino2012} requires the user to write a {\em kernel} that will be run simultaneously by many threads. Threads are grouped into {\em workgroups}, which are run 
on the same vector processor, called \emph{streaming multiprocessor}~(SM). Programmers need to explicitly declare in what memory the data is stored. Further, they can use fast synchronization and communication within a workgroup via {\em local} memory. For \texttt{NVIDIA}~\GPUs, groups of $32$ threads execute in a {\sc simd} fashion; such a group is called {\em warp}.

The main ingredients for an efficient many-core implementation are (1) exposing sufficient parallelism to fully utilize the device and (2) accessing the memory in an efficient way. The latter one includes techniques that
\begin{enumerate}
    \item[(a)] hide the latency of the memory transfer between host and \GPU, for example by overlapping the kernel execution with the memory transfer, and
    \item[(b)] restructure the program in order to improve both spatial and temporal locality of reference to the global memory of the \GPU.\footnote{Here, \emph{spatial locality} corresponds to the case in which threads in the same warp access consecutive global memory locations in the same instruction. Such a coalesced access is collapsed into a single global memory transaction (in contrast, if threads in a warp access memory with a stride of $32$, then $32$ sequential memory transaction are needed). \emph{Temporal locality} corresponds to the case in which most of the threads in the same workgroup access the same memory location in the same instruction. In this case, the first warp accessing it brings the corresponding data block to the L1 cache within one memory transfer (which is then used by the subsequent warps).}
\end{enumerate}
% Our implementation makes effective use of both ingredients.

\subsection{Massively-Parallel Nearest Neighbor Computations}
%Modern many-core devices such as \GPUs~offer massive parallelism and can nowadays also be used for so-called \emph{general-purpose computations} such as matrix-matrix multiplication. In contrast to standard \CPU-based systems, \GPUs~rely on simplified control units that are better suited for ``simple'' tasks, which can be executed in a massively-parallel manner~\cite{ChengGM2014}. The main ingredients for an efficient many-core implementation are (1) exposing sufficient parallelism to the device and (2) accessing the memory in an efficient way (\eg, data transfer between host and device). The operations conducted on the \GPU~are based on the \emph{single instruction multiple data}~(SIMD) paradigm, which means that all threads in the same group perform the same operation in a single clock cycle, while possibly accessing different memory locations. \todot{Cosmin: More HPC stuff?}

We address the problem of computing the $k\geq1$ nearest neighbors of all points given in a query set $\queryset=\{\vec{q}_1,\ldots,\vec{q}_\testsize\} \subset \Reals^\tdim$ w.r.t. to all points provided in a reference set $\pointset=\{\vec{x}_1,\ldots,\vec{x}_\tsize\} \subset \Reals^\tdim$. Usually, the ``closeness'' between two points is defined via the Euclidean distance (which we will use), but other distance measures can be applied as well. Such nearest neighbor computations form the basis for a variety of methods both in data mining and machine learning including proximity-based outlier detection, classification, regression, density estimation, and dimensionality reduction, see, \eg, Hastie~\etal~\cite{HastieTF2009}. The task of computing the induced distances (and keeping track of the list of neighbors per object) can be addressed naively in a brute-force manner spending $\O(\tsize \testsize \cdot (\tdim + \log k))$ time, which quickly becomes computationally very demanding. Massively-parallel computing can significantly reduce the runtime in this case as shown by Garcia~\etal~\cite{GarciaDNB2010}. Still, for large query and reference sets, the computational requirements can become very large.

Various other approaches have been proposed in the literature that aim at taking advantage of the computational resources provided by \GPUs~in combination with other techniques~\cite{BustosDHK2006,PanM2011,SismanisPS2012}. The focus of this work is on massively-parallel processing of \kdtrees. While several implementations have been proposed that address such traversals from a more general perspective (\eg, in the context of \emph{ray tracing})~\cite{HeinermannKPG2013,Nakasato2012,QiuMN2009,WangC2010}, these approaches are not suited for nearest neighbor search in moderate-sized feature spaces (\ie, $\tdim>3$), except for the recently proposed \bufferkdtree~extension~\cite{GiesekeHOI2014}.

% \todot{Fabian, for completeness would it be worth to give the average O time for \bufferkdtree extension?}

\subsection{Nearest Neighbor Search via \KdTrees}
%As mentioned above, 
Spatial search structures such as \kdtrees can be used to speed up nearest neighbor search. \Kdtrees can be constructed as follows~\cite{Bentley1975,FriedmanBF1977}: For a given point set $\pointset$, a \kdtree is a binary tree with the root corresponding to $\pointset$. The children of the root are obtained by splitting the point set into two (almost) equal-sized subsets, which are processed recursively. In their original form, \kdtrees are obtained by resorting to the median values in dimension $i \bmod \tdim$ to split a point set corresponding to a node $v$ at level $i$ (starting with the root at level $0$).\footnote{Other splitting rules might be applied (\eg, according to the ``longest'' side).} The recursive process stops as soon as a predefined number of points are left in a subset. The \kdtree stores the splitting values in its internal nodes; the points corresponding to the remaining sets are stored in the leaves.

The tree structure can be used to accelerate nearest neighbor search: Let $\vec{q} \in \Reals^\tdim$ be a query point. For the sake of exposition, we focus on $k=1$ (the case of $k>1$ neighbors works similarly). The overall search takes place in two phases. In the first one, the tree is traversed from top to bottom to find the $\tdim$-dimensional cell (induced by the tree/splitting process) that contains the query point~$\vec{q}$. Going down the tree can be conducted efficiently using the median values stored in the internal nodes of the \kdtree. In the second phase, the tree is traversed bottom-up, and on the way back to the root, subtrees are checked in case the query point is close to the corresponding splitting hyperplane. If the distance of the query point $\vec{q}$ to the hyperplane is less than the distance to the current nearest neighbor candidate, then the subtree is checked for better candidates (recursively). Otherwise, the whole subtree can be safely pruned (no recursion). Once the root is reached twice, the overall process stops and the final nearest neighbor candidate is returned.

Given a low-dimensional search space, it is usually sufficient to process a relatively small number of leaves, which results in a logarithmic runtime behavior (\ie, $\O(\log \tsize)$ time per query in practice). However, the performance usually decreases for increasing $\tdim$ due to the curse of dimensionality. In the worst case, all nodes and leaves of the \kdtree need to be processed, which yields a linear query time (\ie, $\O(\tsize)$ time per query).

\subsection{Revisited: \BufferkdTrees}

\begin{wrapfigure}{r}{0.42\textwidth}
\vspace{-62pt}
\centering{
\resizebox{0.41\textwidth}{!}{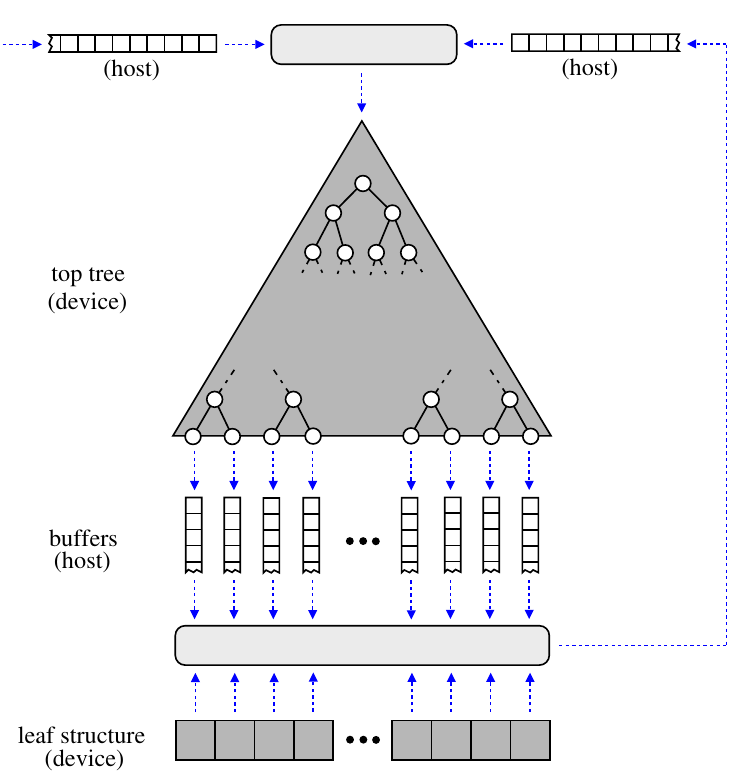}
}
\vspace{-10pt}
\caption{A \bufferkdtree~\cite{GiesekeHOI2014}: The gray elements are stored on \GPU.}
\label{fig:buffer_kdtree}
\vspace{-20pt}
\end{wrapfigure}
A standard multi-threaded \kdtree-based traversal assigns one thread to each query. For \GPUs, such an approach is not suited since each thread might induce a completely different path, which significantly shortens their computational benefits. The main idea of the \bufferkdtree extension is to delay the processing of the queries by buffering similar patterns prior to their common processing~\cite{GiesekeHOI2014}. The reorganized workflow is based on \bufferkdtrees, which are sketched next, followed by a description of the buffered nearest neighbor search.

A \bufferkdtree consists of (1) a top tree, (2) a leaf structure, (3) a set of buffers, and (4) two input queues that store the queries, see Figure~\ref{fig:buffer_kdtree}. The top tree corresponds to a classical \kdtree with its splitting values (\eg, medians) laid out in memory in a pointer-less manner. The leaf structure stores the point sets that stem from the splitting process in a consecutive manner. In addition, a buffer is attached to each leaf of the top tree that can store $B$ query indices (\eg, $B=1024$). The input queues are used to store the input query indices and the query indices that need further processing after a \texttt{ProcessAllBuffers} call.

\begin{algorithm}[t]
\caption{\textsc{LazySearch} \cite{GiesekeHOI2014}}
\label{alg:lazy_search}
\small
\begin{algorithmic}[1]
\REQUIRE A chunk $\queryset=\{\vec{q}_1,\ldots,\vec{q}_\testsize\} \subset \Reals^\tdim$ of query points.
\ENSURE The $k\geq1$ nearest neighbors for each query point.
\STATE Construct buffer \kdtree $\mathcal{T}$ for $\pointset=\{\vec{x}_1,\ldots,\vec{x}_\tsize\} \subset \Reals^\tdim$.
\STATE Initialize queue \texttt{input} with all $\testsize$ query indices.
\WHILE{either \texttt{input} or \texttt{reinsert} is non-empty}
\STATE Fetch $M$ indices $i_1,\ldots,i_M$ from \texttt{reinsert} and \texttt{input}.
\STATE $r_1,\ldots,r_M$ = \textsc{FindLeafBatch}($i_1,\ldots,i_M$)
\FOR{$j=1,\ldots,M$}
\IF{$r_j \neq -1$}
\STATE Insert index $i_j$ in buffer associated with leaf $r_j$.
\ENDIF
\ENDFOR
\IF{at least one buffer is half-full (or queues empty)}
\STATE $l_1,\ldots,l_N$ = \textsc{ProcessAllBuffers}()
\STATE Insert $l_1,\ldots,l_N$ into \texttt{reinsert}.
\ENDIF
\ENDWHILE
\STATE \textbf{return} list of $k$ nearest neighbors for each query point.
\end{algorithmic}%
\end{algorithm}%

% \subsubsection{Workflow:}
A \bufferkdtree~can be used to delay the processing of the queries by performing several iterations, see Algorithm~\ref{alg:lazy_search}: In each iteration, the procedure \textsc{FindLeafBatch} retrieves indices from both the \texttt{input} and \texttt{reinsert} queue and propagates them through the top tree. The indices, which are stored in the corresponding buffers, are processed in chunks via the procedure \texttt{ProcessAllBuffers} once the buffers get full. All indices that need further processing (\ie, their implicit tree traversal has not reached the root twice) are inserted into \texttt{reinsert} again. Thus, in each iteration, one (1) finds the leaves that need to be processed next and (2) updates the queries' nearest neighbors. 

While the first phase is not well-suited for massively-parallel processing, the second one is and, since it constitutes the most significant part of the runtime, yields valuable overall speed-ups. The main advantage of the reorganized workflow is that all queries are processed in the same block-wide {\sc simd} instruction and exhibit either good spatial or temporal locality of reference, \ie, coalesced or cached global memory accesses (see Section~\ref{subsec:GPUarch}).
%with consecutive threads accessing consecutive memory locations. 
For details of the particular many-core implementation of the procedure \textsc{ProcessAllBuffers}, we refer to Gieseke~\etal~\cite{GiesekeHOI2014}. Note that both the top tree and the leaf structure need to be stored on the \GPU~given the original implementation~\cite{GiesekeHOI2014}---this limits the amount of reference patterns that can be processed. 
% Next, we show how to remove this limitation without decreasing the performance of the massively-parallel search.

%  The resulting many-core implementation can be seen as an intermediate version between a standard \kdtree based search and a massively parallel brute-force approach.

\section{Processing Bigger Trees}
\label{sec:processing_bigger_trees}
One issue not addressed so far is the fact that the memory of modern \GPUs~is still relatively small compared to host memory.\footnote{We assume that the host's memory is large enough to store the whole \bufferkdtree.} This limits the amount of data points that can be processed. We now describe modifications that allow the \bufferkdtrees~to scale to massive data sets not fitting on a \GPU~anymore.

\subsection{Construction Phase}
As shown below, one can basically process arbitrarily large query sets by considering chunks of data points. Dealing with huge reference sets, however, is more difficult: Since the top tree and the full leaf structure (that stores the rearranged reference points) have to be made available to all threads during the execution of \textsc{ProcessAllBuffers}, one cannot directly split up the leaf structure. However, as explained next, one can avoid storing the leaf structure in its full entirety on the many-core device without significantly increasing the overall runtime.

\begin{wrapfigure}{r}{0.5\textwidth}
\vspace{-18pt}
\centering{
\resizebox{0.48\textwidth}{!}{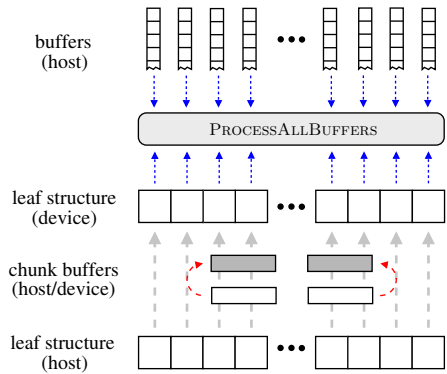}
}
% \vspace{-10pt}
\caption{Adapted memory layout: The leaf structure is \emph{not} stored explicitly on the device. Instead, memory for two chunk buffers is allocated on both the device and the host, which is used for an concurrent compute-and-copy processing of the leaf structure.}
\label{fig:memory_layout}
\vspace{-18pt}
\end{wrapfigure}
% Both the top tree and the leaf structure have to be stored on the device (in its original form). 
We start by focusing on the space needed for the top tree: From a practical perspective, a small top tree is usually advantageous compared to the full tree used by a classical \kdtree-based search, since the efficiency gain on \GPUs~stems from processing ``big'' leaves. For instance, given a reference set $\pointset \subset \Reals^{10}$ with two million points, top trees of height $h=8$ or \mbox{$h=9$} are usually optimal~\cite{GiesekeHOI2014}. Further, since only median values are stored in the top tree, the space consumption is negligible even given much bigger top trees.\footnote{For a tree of height $h=20$, suitable for more than one billion points, less than ten megabytes are needed. Note that the space for the buffers (\eg, of size 128 each) stored on the host does usually not cause any problems (\eg, less then a gigabyte).} 
% removed by FABIAN
% Hence, since spatial search structures are, in general, only suited for relatively low-dimensional search spaces (\eg, up to $\tdim=30$), the space needed by the top tree on the many-core device does not cause any problems.
% Note that the additional space needed for the top tree and the buffers is very small compared to the leaf structure due to the fact that a small tree height is usually enough. 
% For instance, for a reference set of size one billion in a ten dimensional feature space, an appropriate top tree of height $h=18$ accommodates less than a megabyte of space; further, the associated buffers (of, \eg, size 256) accommodate about 200 megabyte. 
% The construction time for a \bufferkdtree is usually quite small compared to the one needed for processing the queries. This is due to the fact that relatively small top trees are usually sufficient from a practical perspective (see above). 
In addition, the top tree can be built efficiently via linear-time median finding~\cite{BlumFPRT1973}, which results in $\O(h \log \tsize)$ time for the whole construction phase. 
% Again, as argued above, top trees are usually much smaller compared to standard \kdtrees---the computation time is, thus, relatively small compared to the subsequent querying phase. 
Hence, one can~(1) build the top tree efficiently on the host system and (2) store it in its full entirety on the \GPU.

% Hence, since a quite large amount of query patterns is given for the scenarios addressed by this work, the construction of the top tree (and, thus, the whole \bufferkdtree) can be conducted sequentially on the host system.

The main space bottleneck stems from the leaf structure.
% which stores the reference points in a rearranged manner. 
For instance, one billion points in $\Reals^{15}$ occupy about 60 gigabytes of space---too much for a modern many-core device. For this reason, we do \emph{not} copy the leaf structure from the host to the device after the construction of the top tree. Instead, we allocate space for two chunk buffers of fixed size on the many-core device. These chunks will be used to overlap the execution of the \textsc{ProcessAllBuffers} procedure with the host-to-device memory transfer for the next chunk. 
%These chunks will be used to conduct an overlapping compute-and-copy process each time the procedure \textsc{ProcessAllBuffers} is invoked. 
Note that we also allocate two associated memory buffers on the host (\emph{pinned memory}~\cite{Scarpino2012}) to achieve efficient concurrent compute and copy operations. The overall memory layout is shown in Figure~\ref{fig:memory_layout}, where the memory that is actually allocated on the device is sketched via gray rectangles.

\subsection{Query Phase}
We now describe the details of the modified querying process. The idea is to keep the leaf structure on the host system and to process the buffers via chunks with concurrent compute and copy operations.

\vspace{-0.2cm}
\subsubsection{Processing the Leaf Structure:}
In each iteration of Algorithm~\ref{alg:lazy_search}, the procedure \textsc{ProcessAllBuffers} is invoked to retrieve all query indices from the buffers attached to the leaves. The queries are processed in a massively-parallel fashion, where each thread compares a particular query with all reference points stored in the associated leaf. Given the modified memory layout, one can now process the leaves in chunks in the following way: The leaf structure containing all~$\tsize$ rearranged reference points is split into $1 < \numleafchunks \ll \tsize$ chunks $C_1, \ldots, C_\numleafchunks$ (\eg, $\numleafchunks=10$). Each chunk~$C_j$ contains the points of the leaf structure at positions $k =C_j^L, \ldots, C_j^R$, where $C_j^L= \ceil{\frac{(j-1) \cdot \tsize}{N}}$ and $C_j^R=\ceil{\frac{j \cdot \tsize}{\numleafchunks}}$. A buffer attached to the top tree corresponds to a leaf in the leaf structure with leaf bounds $0\leq l_i<r_i \leq \tsize-1$. All queries removed from the buffers are then processed in $\numleafchunks$ iterations and a query $i$ with leaf bounds $l_i$ and $r_i$ is processed in iteration $j \in \{1,\ldots,\numleafchunks\}$ if $[l_i,r_i] \cap [C_j^L, C_j^R] \neq \emptyset$, \ie, if the leaf bounds overlap with chunk $C_j$.

The chunks are processed sequentially $C_1, C_2, \ldots, C_\numleafchunks$. The data needed for each chunk is copied from the host to one of the two chunk buffers allocated on the device prior to conducting the brute-force computations, see Figure~\ref{fig:memory_layout}. To hide the overhead induced for these copy operations, the copy process for the next chunk is started as soon as the computations for its predecessor have been invoked. In particular, the processing of a chunk $C_j$ takes place in three phases: 
\begin{enumerate}
 \item[(1)] \emph{Brute:} First, the massively-parallel brute-force nearest neighbor computations are invoked (non-blocking kernel call). The data needed for these computations (chunk $C_j$) have been copied in round $j-1$ (for $C_0$, the data is either available from an initial copy operation or from the previous round).
 \item[(2)] \emph{Copy:} While the brute-force computations are conducted by the \GPU, the data for the next chunk are copied from host to the buffer on the device that is currently not in use.\footnote{For $j=\numleafchunks$, the data for chunk 0 are copied from host to the corresponding buffer on the device for next round (\ie, next call of \textsc{ProcessAllBuffers}).} Note that copy operations on the host system have to be conducted as well to ensure that the correct part of the leaf structure is moved to the appropriate pinned memory buffer (which is then copied to the device), see again Figure~\ref{fig:memory_layout}.
 \item[(3)] \emph{Wait:} In the final phase, one simply waits for the kernel invoked in the first phase to finish its computations (blocking call).
\end{enumerate}

The iterative compute-and-copy processing of the chunks can be implemented via two (\texttt{OpenCL}) command queues~\cite{Scarpino2012}: For the first chunk, phase (1) and (3) are instantiated via command queue A, whereas the copy process (2) is instantiated via command queue B (non-blocking for both (1) and (2)). For the second chunk, phases (1) and (3) are instantiated via command queue B and (2) via command queue A. This process continues until all chunks have been processed. In essence, the use of two command queues allows the copy phase (2) to run in parallel with the brute-force computation phases (1) and (3).

% Note that only memory for the two chunk buffers needs to be allocated on the device and the potential overhead for copying leaf chunks from host to device can be hidden (see Section~\ref{sec:experiments}). 
Given the new workflow, one can basically handle an arbitrary amount of reference patterns. The only restriction is the memory available on the host. In case not enough main memory is available, one can store the leaf structure on disk and copy the chunks from disk to device memory (via host memory).\footnote{Depending on the particular architecture, the induced copying processes might become a major bottleneck. However, one can shorten this drawback by increasing the leaf size of the \bufferkdtree~such that more computations have to be conducted for each transfer of data from disk to device.}

% However, since a quite large amount work is usually done in the leaves, the overhead for copying the data is usually completely hidden.

\subsubsection{Multi-Many-Core Querying:}
Assuming that a fixed amount of memory is available on the many-core device to store the query patterns and the results at all times (\eg, one gigabyte), one can process an arbitrarily large query set by removing fully processed indices and by adding new indices on-the-fly, see Algorithm~\ref{alg:lazy_search}. An even simpler approach is to split up the queries into chunks and to handle these chunks independently. One drawback of the latter approach could be the overhead induced by applying the procedure \texttt{ProcessAllBuffers} in case the buffers are not sufficiently filled (which usually takes place as soon as no queries are available anymore). However, given relatively large chunks, the induced overhead is very small as shown in our experimental evaluation. 

In a similar fashion, one can make use of multiple many-core devices by splitting all queries into ``big'' chunks according to the devices that are available. These chunks, which might have to be split into smaller chunks as described above, can be processed independently from each other.
% \footnote{The only limitation of this approach is that it is not applicable to scenarios with a huge number of reference points and relatively small query sets.}

\section{Experiments}
\label{sec:experiments}
The purpose of the experiments provided below is to analyze the efficiency of the modified workflow and to sketch the potential of the overall approach in the context of large-scale scenarios. For a detailed experimental comparison including an analysis of the different processing phases and the influence of parameters related to the \bufferkdtree~framework, we refer to our previous work~\cite{GiesekeHOI2014}.

\subsection{Experimental Setup}
All runtime experiments were conducted on a standard desktop computer with an \texttt{Intel(R) Core(TM) i7-4790K} \CPU~running at 4.00GHz (4 cores; 8 hardware threads), 32GB~RAM, and two \texttt{Nvidia GeForce Titan Z}~\GPUs{} (each consisting of two devices with 2880 shader units and 6~GB main memory). The operating system was \texttt{Ubuntu 14.4.3 LTS} (64 Bit) with kernel \texttt{3.13.0-52}, \texttt{CUDA 7.0.65} (graphics driver 340.76), and \texttt{OpenCL 1.2}. All algorithms were implemented in \texttt{C} and \texttt{OpenCL}, where \texttt{Swig} was used to obtain appropriate \texttt{Python} interfaces.\footnote{The code is publicly available under \url{https://github.com/gieseke/bufferkdtree}.} The code was compiled using \texttt{gcc-4.8.4} at optimization level \texttt{-O3}.

For the experimental evaluation, we report runtimes for both the construction and the query phase (referred to as ``train'' and ``test'' phases), where the focus is on the latter one (that makes use of the \GPUs). We consider the following three implementations:

\begin{enumerate}\itemsep0pt\parskip0pt
 \item[(1)] \texttt{bufferkdtree(i)}: The adapted \bufferkdtree~implementation with both \texttt{FindLeafBatch} and \texttt{ProcessAllBuffers} being conducted on $i$ \GPUs.
%  \item[(2)] \texttt{bufferkdtree(cpu)}: a corresponding sequential \CPU~variant that only operates on the host system.
 \item[(2)] \texttt{kdtree(i)}: A multi-core implementation of a \kdtree-based search, which runs $i$ threads in parallel on the \CPU~(each handling a single query).
%  \item[(4)] \texttt{kdtree(gpu)}: a na\"{\i}ve \GPU~implementation, which traverses an appropriately built \kdtree in parallel (one thread per query).
 \item[(3)] \texttt{brute(i)}: A brute-force implementation that makes use of $i$~\GPUs~to process the queries in a massively-parallel manner.
\end{enumerate}

The parameters for the \bufferkdtree implementation were fixed to appropriate values.\footnote{For a tree of height $h$, we fixed $B=2^{24-h}$ and the number $M$ of indices fetched from \texttt{input} and \texttt{reinsert} in each iteration of Algorithm~\ref{alg:lazy_search} to $M=10\cdot B$. Note that the particular assignments for these and other parameters did not have a significant influence on the performance as long as they were set to reasonable values.} Note that both competitors of \texttt{bufferkdtree} have been evaluated extensively in the literature; the reported runtimes and speed-ups can thus be put in a broad context. For simplicity, we fix the number $k$ of nearest neighbors to $k=10$ for all experiments.

We focus on several data-intensive tasks from the field of astronomy. Note that a similar runtime behavior can be observed on data sets from other domains as well as long as the dimensionality of the search space is moderate (\eg, from $\tdim=5$ to $\tdim=30$). We follow our previous work and consider the \texttt{psf\_mag}, \texttt{psd\_model\_mag}, and \texttt{all\_mag} data sets of dimensionality $\tdim=5$, $\tdim=10$, and $\tdim=15$, respectively; for a description, we refer to Gieseke~\etal~\cite{GiesekeHOI2014}. In addition, we consider a new dataset derived from the \emph{Catalina Realtime Transient Survey} (\texttt{crts})~\cite{DrakeEtAL2009,DjorgovskiEtAl2011,MahabalDDDGWCMTBL2011}. This survey contains tens to hundreds of observations for more than 500 million sources over a large part of the sky. The resulting light-curves (time-series of light received as a function of time) are used to derive several statistical features.\footnote{In particular, we make use of the $amplitude$, $Stetson_j$, $Stetson_k$, $Skew$, $fpr_{mid35}$, $fpr_{mid50}$, $fpr_{mid65}$, $fpr_{mid80}$, $shov$, $maxdiff$~\cite{RichardsEtAl2011,FarawayMSWYZ2014}.} 
We use ten such features on a set of 30 million light-curves for the experiments described here. The main interest is to find outliers in the large space that can lead to interesting discoveries. 
% That will be covered in a later paper.
% from so-called light-curves. More precisely, we consider several data set instances of varying dimensionality, which are based on \todot{Ashish: description of the data?}. 

\subsection{Modified Workflow}
While the modified workflow permits the use of \bufferkdtrees for massive data sets that do not fit in the memories of the many-core devices, it might also induce a certain overhead compared to its original version due to the induced copy operations and reduced workload per kernel call. In addition, the ``naive'' use of multiple \GPUs~might exhibit a worse performance compared to filling the \texttt{input} queue with new queries on-the-fly. We now investigate both potential drawbacks.

\subsubsection{Processing Leaves in Chunks:} 
\begin{figure}[t]
\centering
\subfigure[$\tsize=500,000$]{
\resizebox{0.31\columnwidth}{!}{
\includegraphics{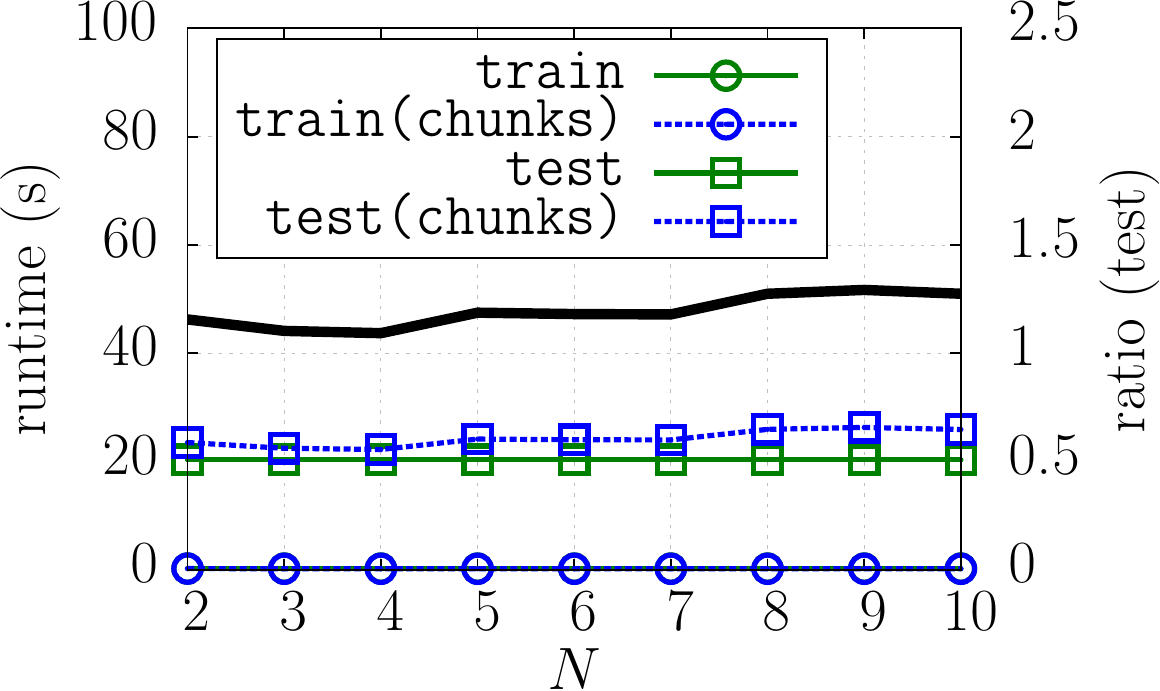}
}
}
\subfigure[$\tsize=1,000,000$]{
\resizebox{0.31\columnwidth}{!}{
\includegraphics{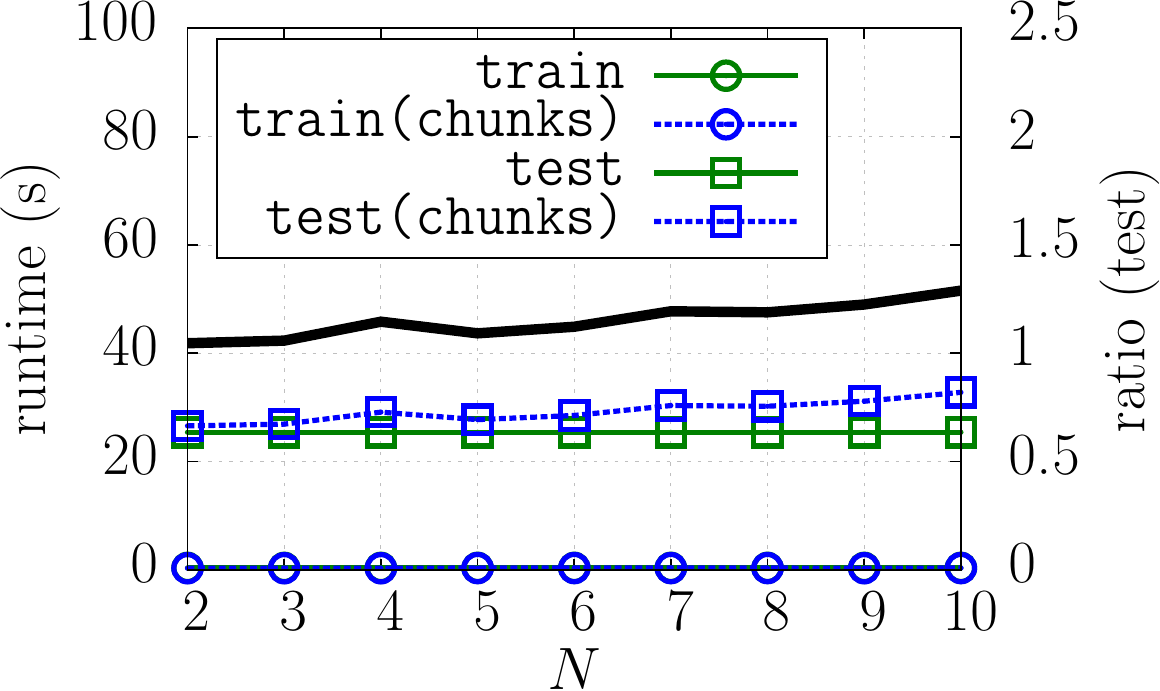}
}
}
\subfigure[$\tsize=2,000,000$]{
\resizebox{0.31\columnwidth}{!}{
\includegraphics{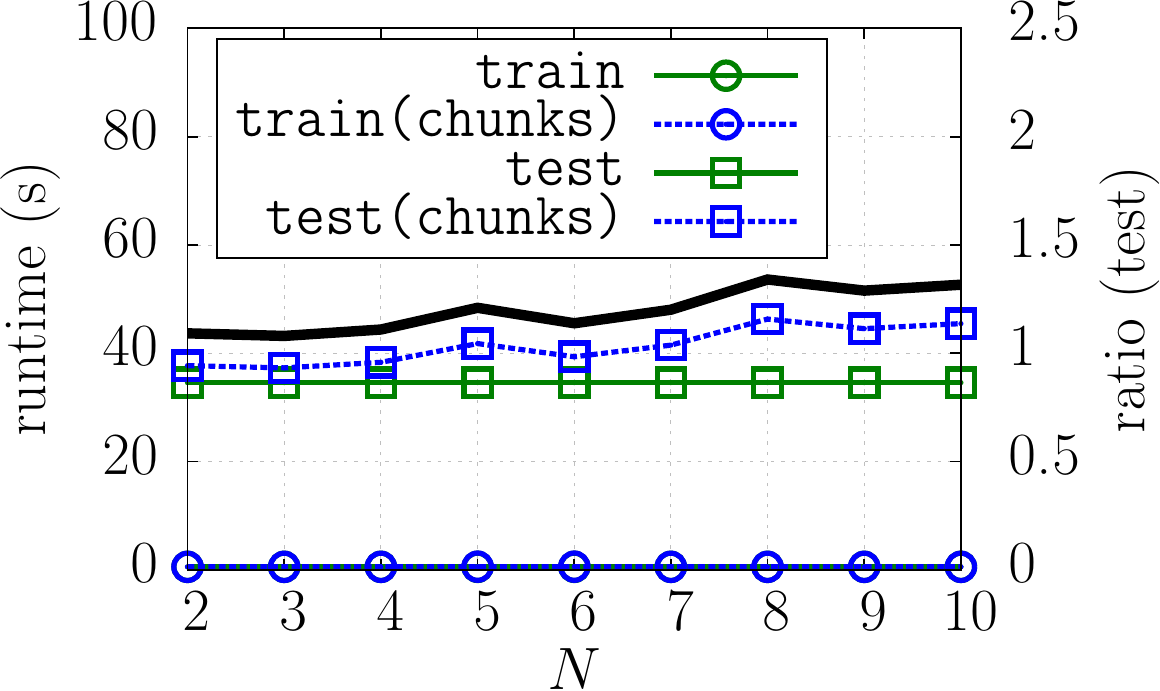}
}
}
\caption{Comparison of the training and testing times between the original workflow (\texttt{train} and \texttt{test}) and the modified workflow that processes the leaves in chunks (\texttt{train(chunks)} and \texttt{test(chunks)}). A varying number $\numleafchunks$ of chunks and training points $\tsize$ are considered using the \texttt{psd\_model\_mag} data set. The number $\testsize$ of test patterns and the tree height $h$ are fixed to $\testsize=10,000,000$ and $h=9$, respectively. The ratio between \texttt{test} and \texttt{test(chunks)} is shown as black, thick line (a value close to 1 indicates a small overhead caused by the chunked processing).}
\label{exps:leaf_chunks}
\end{figure}
The main modification is the different processing of the leaf structure in case it does not fit in the device's memory. To evaluate the potential overhead caused by the additional copy operations (between host and device) during the execution of \textsc{ProcessAllBuffers}, we consider data set instances that still fit in memory and compare the runtimes of~(1) the original workflow with (2) the workflow that is based on multiple chunks. 

In Figure~\ref{exps:leaf_chunks}, the outcome of this comparison is shown for a varying number~$\numleafchunks$ of chunks and a varying number $\tsize$ of training points. In all three cases, the number of test patterns is fixed to $\testsize=10,000,000$. Further, the tree height is set to~$h=9$ and a single \GPU~is used (\ie, \texttt{bufferkdtree(1)}). Two observations can be made: First, the training time is very small compared to the test time for all cases (even though the tree is constructed sequentially on the host). Second, the performance loss induced by the chunked processing is very small for almost any number $\numleafchunks$ of chunks (\ie, the ratio shown as black, thick line is close to 1). In particular, this is the case for smaller values of~$\numleafchunks$; using more chunks naturally yields more overhead, which, however, decreases again if one increases the number of training and test patterns. 

Thus, the runtimes of the new, chunked workflow are close to the one of the original approach---indicating that the overlapping compute-and-copy process successfully hides the additional overhead for the copy operations.

\subsubsection{Multi-Many-Core Processing:} 
As outlined above, one can simply distribute the test queries to multiple devices to take advantage of the additional computational resources. Again, this can lead to a certain overhead, since invoking the procedure \textsc{ProcessAllBuffers} becomes less efficient at the end of the overall processing (and this happens earlier in case the test queries are split into chunks). Similarly to the experiment provided above, we compare the efficiency of \texttt{bufferkdtree(4)} with the one of \texttt{bufferkdtree(1)}, a standard single-device processing with all test patterns fitting on the \GPU. For this sake, we consider $\numleafchunks=1$ leaf chunks (\ie, no modified processing of the leaves), $\tsize=2\cdot {10}^{6}$ training patterns, and vary the number $\testsize$ of test patterns. 

\begin{figure}[t]
\centering
\subfigure[\texttt{psf\_mag}]{
\resizebox{0.3\columnwidth}{!}{
\includegraphics{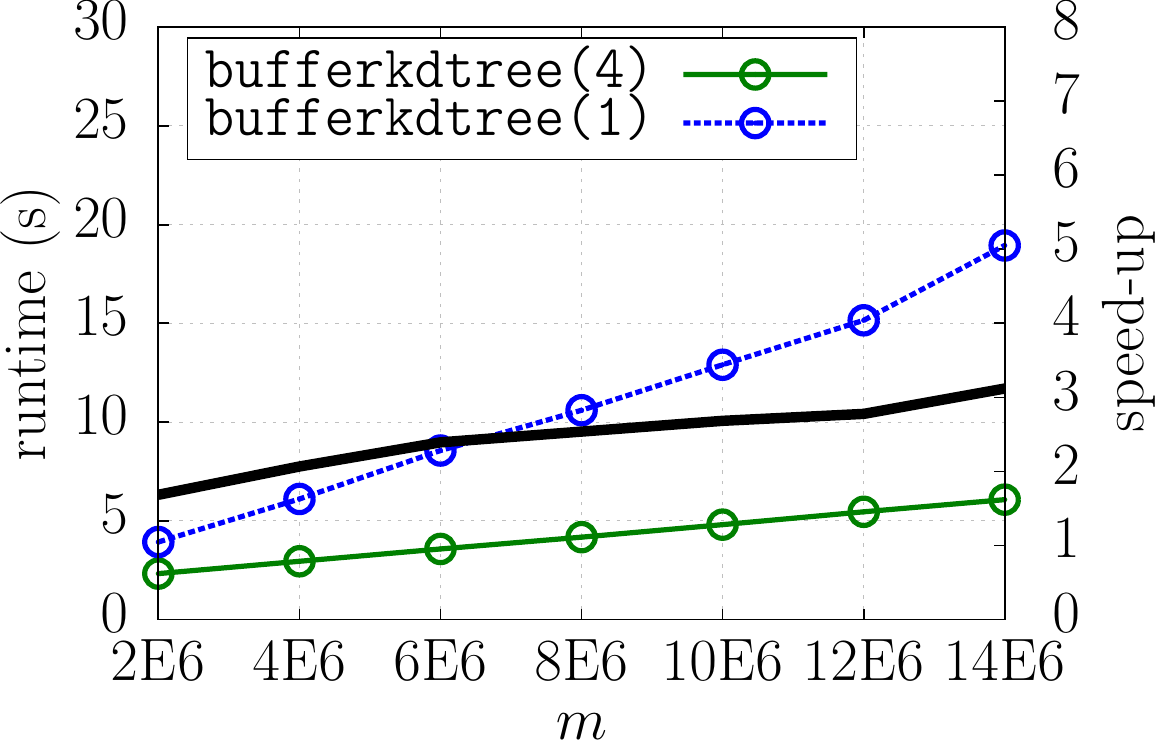}
}
}
\subfigure[\texttt{psf\_model\_mag}]{
\resizebox{0.3\columnwidth}{!}{
\includegraphics{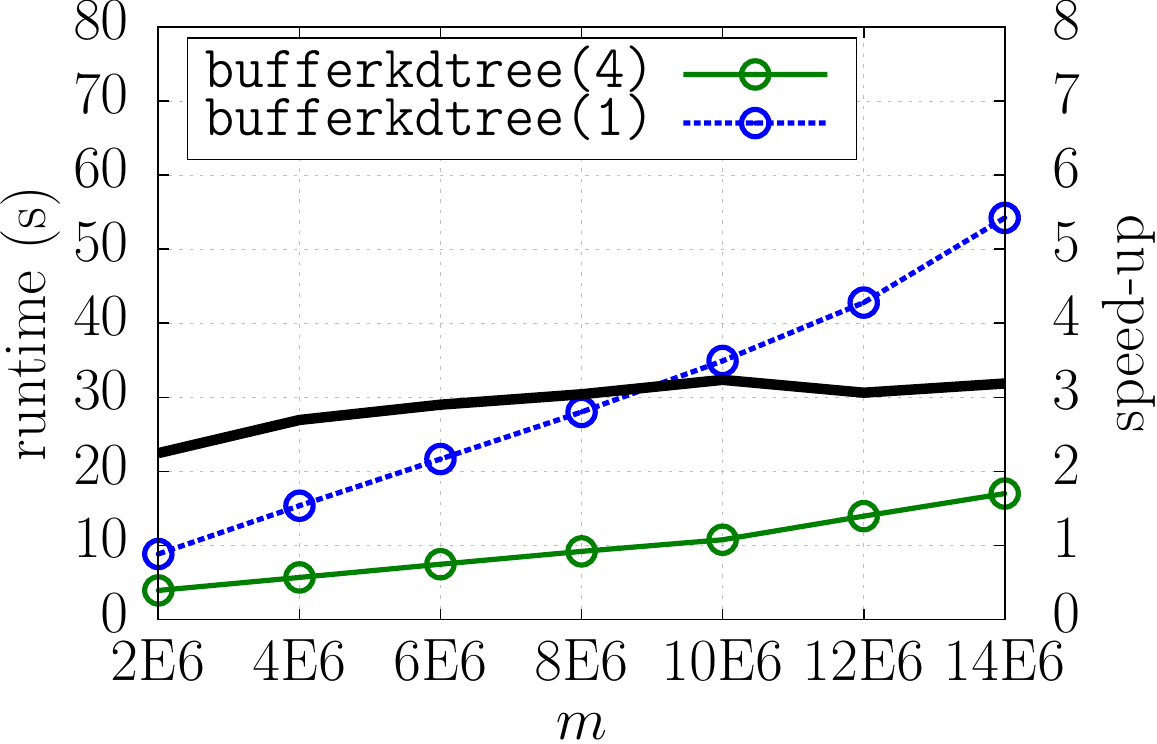}
}
}
\subfigure[\texttt{all\_mag}]{
\resizebox{0.3\columnwidth}{!}{
\includegraphics{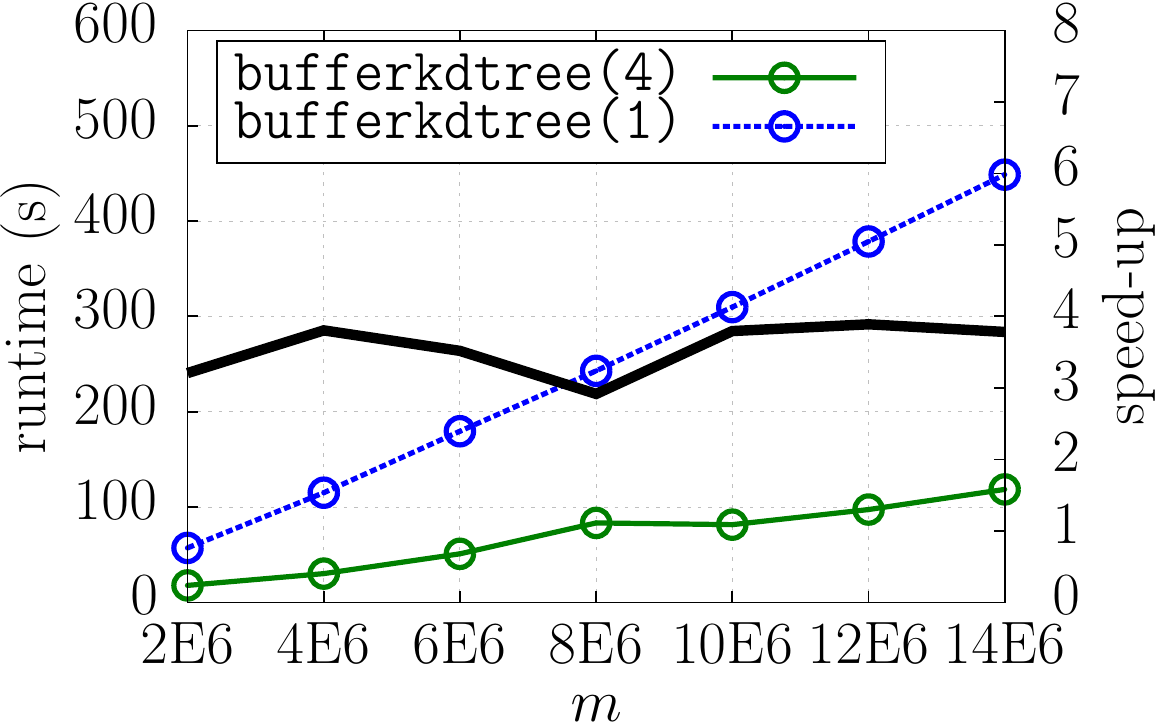}
}
}
\caption{Runtime comparison for the test phase between \texttt{bufferkdtree(1)} and \texttt{bufferkdtree(4)}, where the test queries are distributed uniformly among the devices for the latter one. The speed-up is shown as black, thick line (maximum 4).}
\label{exps:multi_many_core}
\end{figure}
The outcome of this experiment for three different data sets is shown in Figure~\ref{exps:multi_many_core}: It can be seen that a suboptimal speed-up of about 2 is achieved in case a relatively small amount of test patterns is processed. However, as soon as the number of test patterns increases, the speed-up gets closer to 4, which depicts the maximum that can be achieved. Hence, the naive way of using all devices in a given workstation does not yield significant drawbacks as soon as large-scale scenarios are considered, which is the scope of this work.

\subsection{Large-Scale Applications}
% The results described above show that the modified chunked processing of the leaves as well as the multi-many-core processing of the queries are implemented efficiently. 
To demonstrate the potential of the modified framework, we consider two large-scale tasks: (1) the application of nearest neighbor models that are based on very large training sets and (2) large-scale density-based outlier detection. 

\vspace*{-0.3cm}
\subsubsection{Huge Nearest Neighbor Models:}
The first scenario addresses nearest neighbor models~\cite{HastieTF2009} that are based on very large training sets. Such models have been successfully been applied for various tasks in astronomy including the detection of distant galaxies or the estimation of physical parameters~\cite{PolstererZG2013,stensbo-smidt13,kremer:15}.

\begin{figure}[t]
\centering
\subfigure[$\tsize=2 \cdot {10}^6$]{
\resizebox{0.3\columnwidth}{!}{
\includegraphics{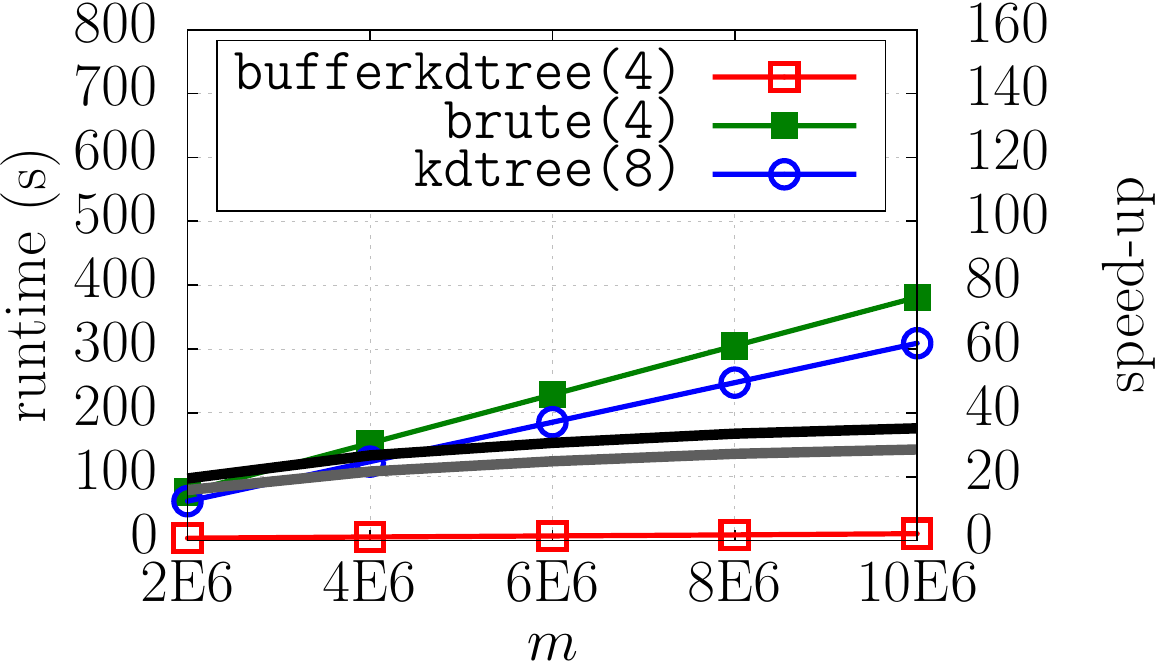}
}
}
\subfigure[$\tsize=4 \cdot {10}^6$]{
\resizebox{0.3\columnwidth}{!}{
\includegraphics{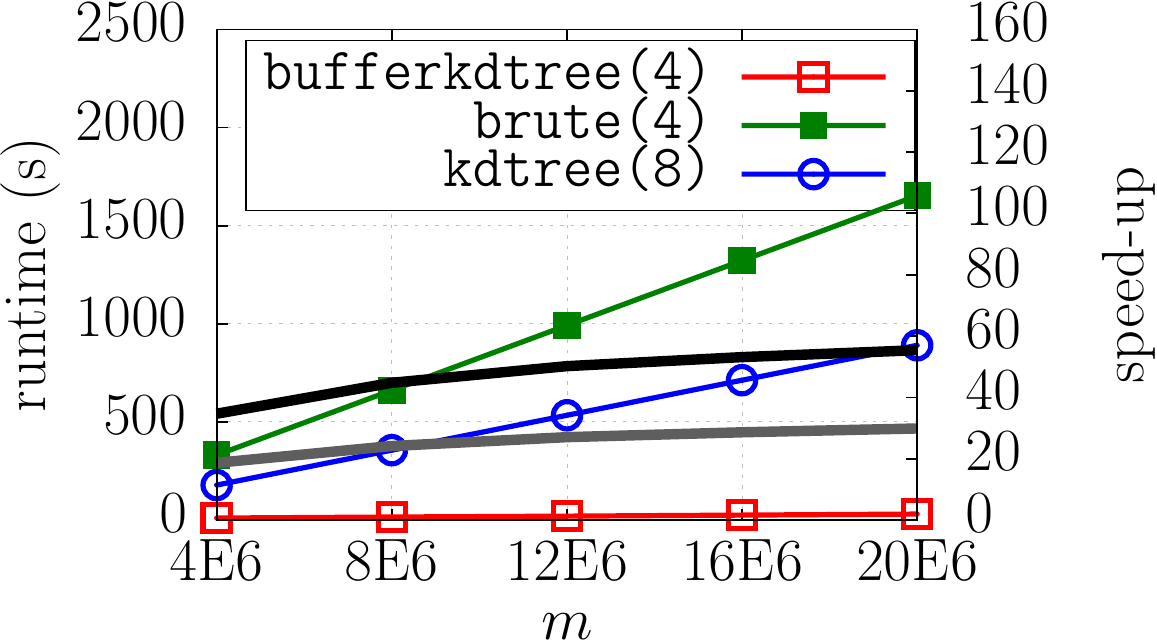}
}
}
\subfigure[$\tsize=6 \cdot {10}^6$]{
\resizebox{0.3\columnwidth}{!}{
\includegraphics{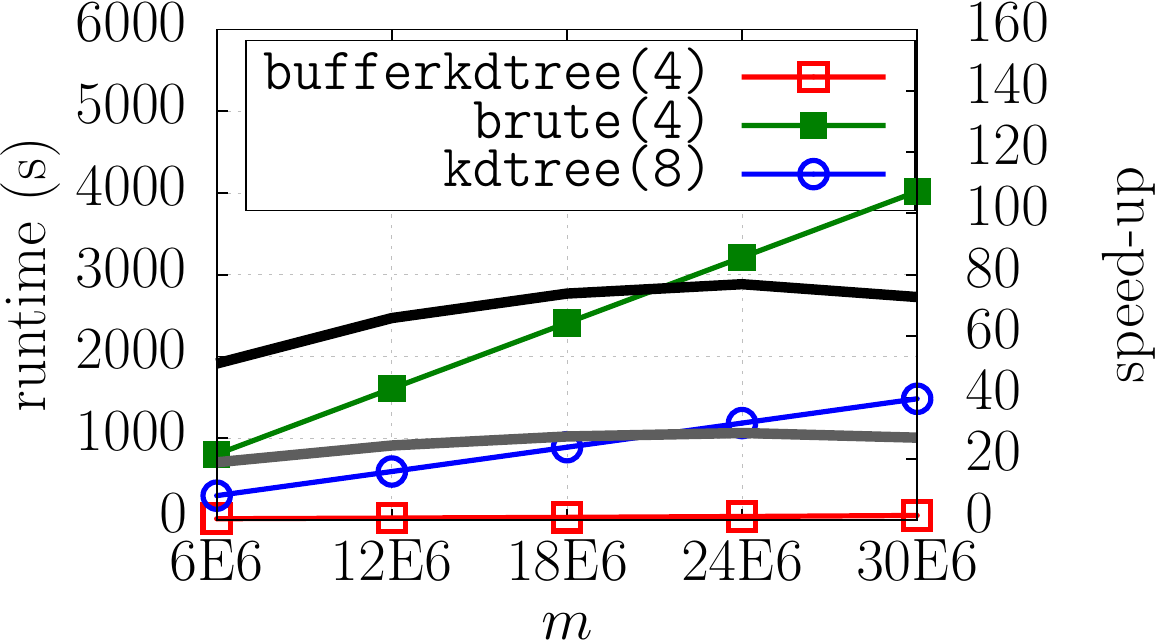}
}
}
\subfigure[$\tsize=8 \cdot {10}^6$]{
\resizebox{0.3\columnwidth}{!}{
\includegraphics{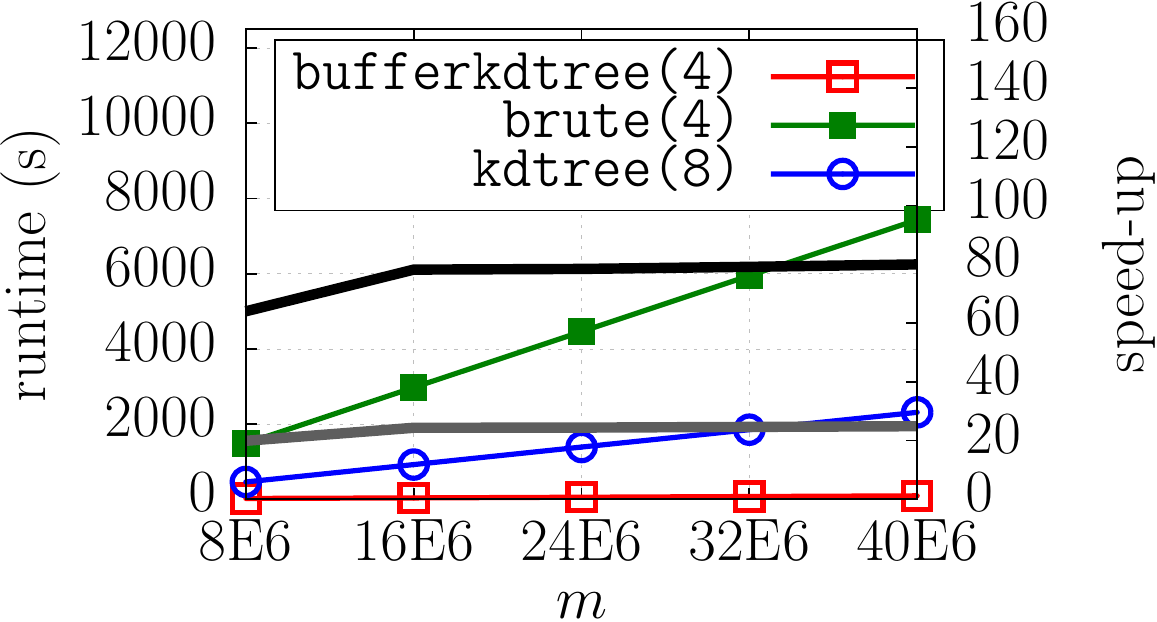}
}
}
\subfigure[$\tsize=10 \cdot {10}^6$]{
\resizebox{0.3\columnwidth}{!}{
\includegraphics{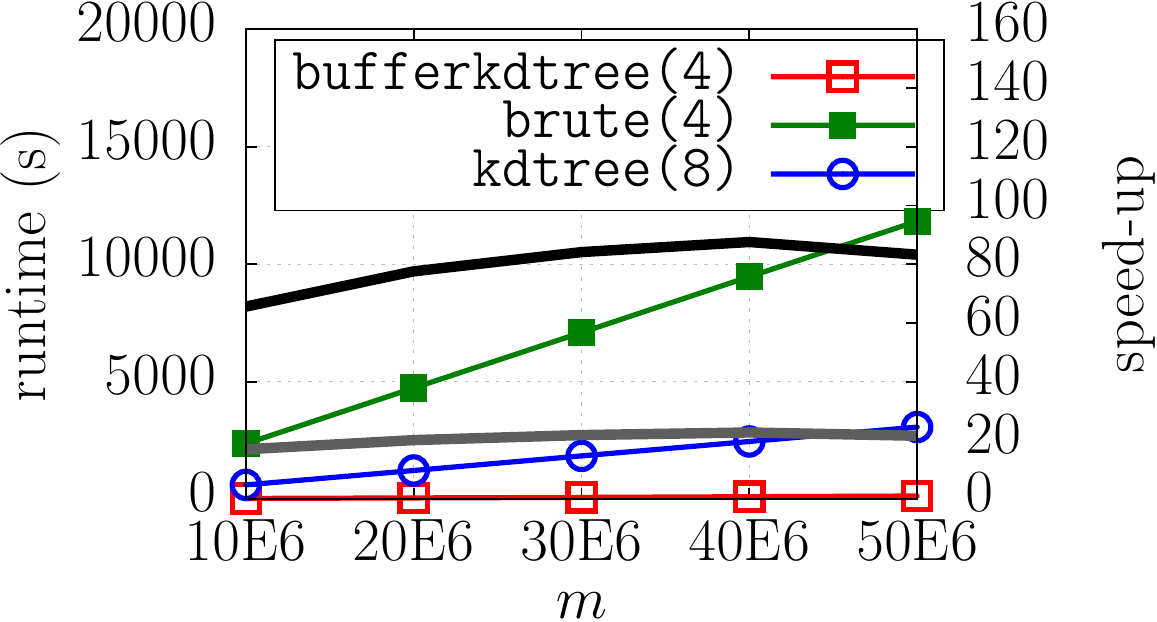}
}
}
\subfigure[$\tsize=12 \cdot {10}^6$]{
\resizebox{0.3\columnwidth}{!}{
\includegraphics{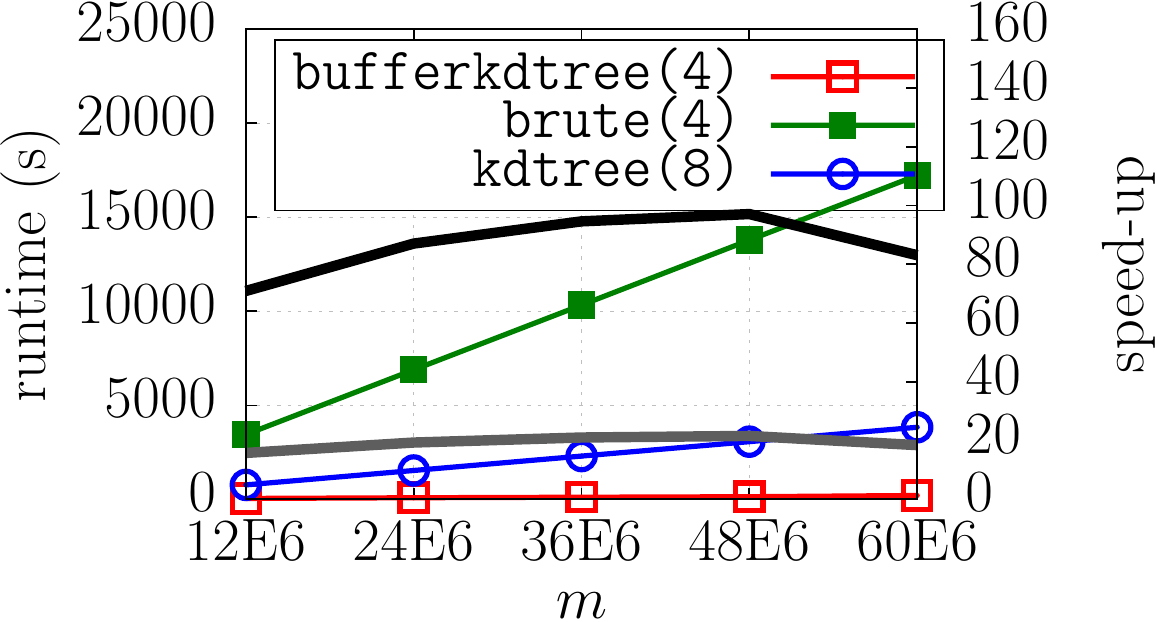}
}
}
\caption{Runtime comparison for a varying number $\tsize$ of training patterns given the \texttt{psf\_model\_mag} data set. The speed-ups of \texttt{bufferkdtree} over \texttt{brute} and \texttt{kdtree} are shown as thick black and gray lines, respectively. In each case, $\testsize=\tsize, \ldots, 5\tsize$ test patterns are considered for the \texttt{bufferkdtree} implementation; for both \texttt{kdtree} and \texttt{brute}, only $\testsize = {10}^6$ were processed to obtain runtime estimates (which are plotted).}
\label{exps:huge_nearest}
\end{figure}
For the experimental comparison we consider scenarios with both a large amount of training and test patterns. More precisely, we consider up to $\tsize=12 \cdot {10}^{6}$ training points and up to $\testsize=5 \cdot \tsize$ test points given the \texttt{psd\_model\_mag} data set. For both tree-based methods, appropriate tree depths are set beforehand (\ie, optimal ones w.r.t. the runtime needed in the test phase). The outcome of this comparison is shown in Figure~\ref{exps:huge_nearest}: It can be seen that valuable speed-ups can be achieved over both competitors. Further, the speed-ups generally become more significant the more patterns are processed. Note that for Figures~(e) and~(f), the \texttt{bufferkdtree} implementation automatically considers $\numleafchunks=3$ chunks (due to the training patterns exceeding the space reserved for them on the device), which results in a slightly worse performance for $\testsize>30 \cdot {10}^6$.

\vspace*{-0.2cm}
\subsubsection{Large-Scale Proximity-Based Outlier Detection:}
As final use case, we consider large-scale proximity-based outlier detection. Various outlier scores have been proposed that are based on the computation of nearest neighbors. A typical one is to rank the points according to their average distance accordiing to their $k$ nearest neighbors, see, \eg, Tan~\etal~\cite{TanSK2005}. Such techniques depict very promising tools in case many reference points are given in a moderate-sized feature space, which is precisely the case for many tasks in astronomy. Typically, these scores require the computation of the nearest neighbors for each of the reference points (known as \emph{all nearest neighbors problem}). Naturally, this can quickly become very time-consuming.

\begin{wrapfigure}{r}{0.5\textwidth}
\vspace{-30pt}
\resizebox{0.45\textwidth}{!}{\includegraphics{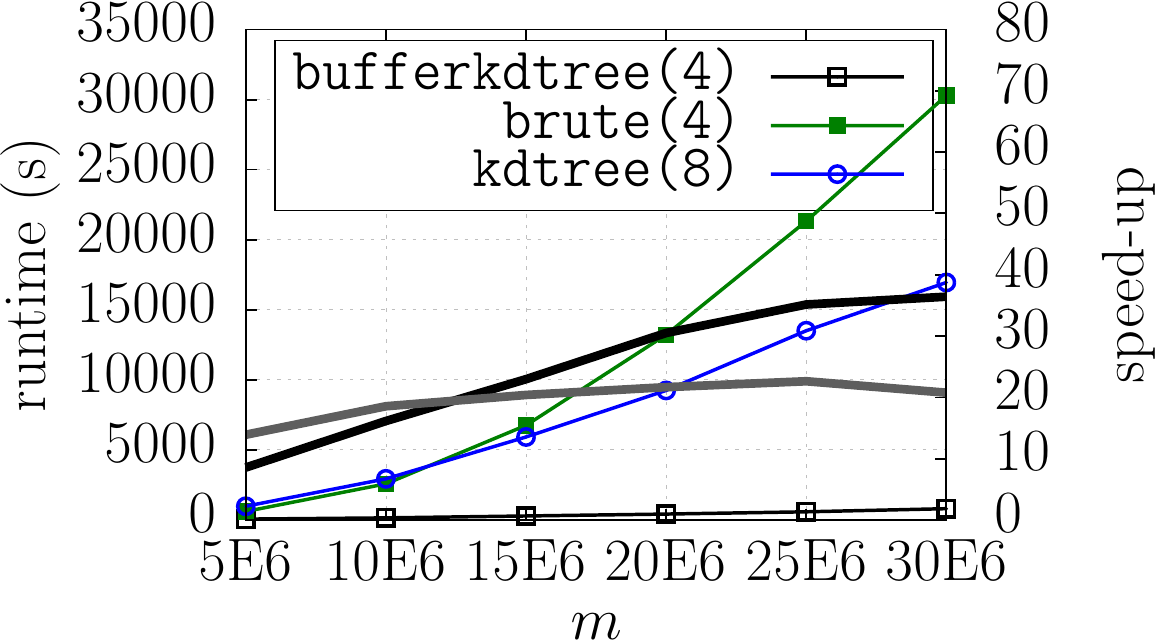}}
\vspace{-5pt}
\caption{Large-Scale Outlier Detection}
\label{fig:speed_ups_outliers}
\vspace{-25pt}
\end{wrapfigure}
To show the potential of our many-core implementation, we consider the \texttt{crts} data set described above (with $\tdim=10$ features) and vary the number $\tsize$ of reference points (here, we have $\tsize=\testsize$ for the full data set). We again compare the performances of all three competitors, where we consider both the runtime for the construction and the one for the query phase. The outcome is shown in Figure~\ref{fig:speed_ups_outliers}. Note that the runtimes for both \texttt{kdtree(cpu,8)} and \texttt{brute(gpu,4)} depict estimates based on a reduced query due to the computational complexity (\ie, up to $\tsize=30\cdot{10}^6$ reference points and a fixed query set of size $\hat{\testsize}=1,000,000$ are considered; the runtime estimates w.r.t. to the full data set instances are plotted). For the \texttt{bufferkdtree(gpu,4)} implementation, we fix $\numleafchunks=3$. It can be seen that the \bufferkdtree implementation yields valuable speed-ups and can successfully process the whole data set in a reasonable amount of time.

% 
% % ****************************************************************************
% % BIBLIOGRAPHY AREA
% % ****************************************************************************
\section{Conclusions}
\label{sec:conclusions}
We provide a modified workflow for processing huge amounts of nearest neighbor queries using \bufferkdtrees. The key idea is to process both the reference and the query points in chunks. While the latter is relatively easy to implement (even given several many-core devices), processing the reference points in chunks is more difficult. As shown in our work, one can effectively hide the overhead induced by the chunked processing of the reference points by interleaving the compute and copy operations. The experiments conducted on commodity hardware demonstrate that a single workstation is enough to efficiently process millions of reference and query points. Future work might address scenarios that are based on even larger reference sets (\eg, hundreds of billions of points), which could also necessitate the efficient construction of the \bufferkdtree. In addition, similar (chunked) buffering techniques might be useful to achieve efficient massively-parallel implementations for other techniques as well.
% (due to a large but comparable small query set).

\paragraph{Acknowledgements.}
The authors would like to thank the \emph{Radboud Excellence
  Initiative} of the Radboud University Nijmegen and \emph{Nvidia} for
its support and generous hardware donations (FG), the \emph{Danish
  Industry Foundation} through the \emph{Industrial Data Analysis
  Service} (CI, CO), the \emph{The Danish Council for
Independent Research\,$|$\,Natural Sciences} through the project \emph{Surveying the
sky using machine learning} (CI), and ACP, IUCAA, IUSSTF, and NSF (AM).

\begin{footnotesize}

%\bibliographystyle{splncs03}
%\bibliography{biblio}

\end{footnotesize}

% ****************************************************************************
% END OF BIBLIOGRAPHY AREA
% ****************************************************************************

\end{document}